# Multiple Charge Density Wave States at the Surface of TbTe$_3$

Ling Fu[1], Aaron M. Kraft[1], Bishnu Sharma[1], Manoj Singh[1], Philip Walmsley[2,3], Ian R. Fisher[2,3], Michael C. Boyer[1*]

[1]Department of Physics, Clark University, Worcester, MA 01610, USA

[2]Geballe Laboratory for Advanced Materials and Department of Applied Physics, Stanford University, Stanford, California 94305-4045, USA

[3]Stanford Institute for Materials and Energy Sciences, SLAC National Accelerator Laboratory, 2575 Sand Hill Road, Menlo Park, California 94025, USA

**Abstract:**

We studied TbTe$_3$ using scanning tunneling microscopy (STM) in the temperature range of 298 – 355 K. As seen in previous STM measurements on RTe$_3$ compounds, our measurements detect a unidirectional charge density wave state (CDW) in the surface Te-layer with a wavevector consistent with that of the bulk, $\boldsymbol{q}_{cdw} = 0.30 \pm 0.01 \boldsymbol{c}^*$. However, unlike previous STM measurements, and differing from measurements probing the bulk, we detect two perpendicular orientations for the unidirectional CDWs with no directional preference for the in-plane crystal axes (*a*- or *c*-axis) and no noticeable difference in wavevector magnitude. In addition, we find regions in which the bidirectional CDW states coexist. We propose that observation of two CDW states indicates a decoupling of the surface Te-layer from the rare-earth block layer below, and that strain variations in the Te surface layer drive the local CDW direction to the specific unidirectional or, in rare occurrences, bidirectional CDW orders observed. This indicates that similar driving mechanisms for CDW formation in the bulk, where anisotropic lattice strain energy is important, are at play at the surface. In our bias-dependent measurements, we find no contrast inversion for the CDW state between occupied and empty states. This finding differs from other quasi 2-dimensional materials containing a "hidden" 1-dimensional character which leads to a favorable Fermi surface nesting scenario. Our temperature-dependent measurements provide evidence for localized CDW formation above the bulk transition temperature, $T_{cdw}$.

**Introduction:**

Charge density wave (CDW) states are broken-symmetry states prevalent in condensed matter systems, where they are often found to coexist and/or compete with other orders. In particular, in the intensely-studied graphene-related systems [1], layered chalcogenides [2,3], organic compounds [4], cuprate high-temperature superconductors [5-9], and BiS$_2$ superconductors [10], CDW states are observed to coexist with superconductivity. However, the specifics of how CDWs interact with other orders, on the nanoscale, are far from understood. Furthermore, even a fundamental understanding of CDW states and their origin remains incomplete.

---

* To whom correspondence should be addressed: mboyer@clarku.edu

While CDW states in some materials are believed to be understood within the more traditional theoretical framework of electronic instabilities driven by Fermi-surface nesting [11-14], often referred to as a Peierls instability, recent experimental and theoretical work shows that CDW states in other material systems are not understood within this framework. Instead, the properties of CDW states in these systems, including members of the transition-metal dichalcogenide family, are determined by strongly momentum-dependent electron-phonon coupling.[15-19] Developing an understanding of CDW states and their origins is critical to gaining insight into complex systems where a CDW state may be only one of several orders present.

The rare-earth tritellurides ($RTe_3$ where R = rare-earth ion) are compounds where the understanding of the driving mechanism for CDW states has evolved. The $RTe_3$ compounds are quasi 2-dimensional materials which have an alternating structure of neighboring conducting square Te planes (double-Te planes) separated by insulating rare-earth block layers, as seen in Figure 1a. Below $T_{cdw}$, a bulk unidirectional incommensurate CDW state is established primarily within the Te planes along the *c*-crystal axis with $q_{cdw} \sim (2/7)c^*$.[20-23] The $RTe_3$ compounds with heavier rare-earth ions (R = Dy, Ho, Er, Tm, and Tb) undergo a second CDW transition at lower temperatures where a CDW state along the *a*-axis, $q_{cdw2} \sim (1/3)a^*$, is established coexisting with the *c*-axis CDW.[24,25]

Angle-resolved photoemission spectroscopy measurements coupled with electronic structure calculations show that the Fermi surface of the $RTe_3$ compounds have significant parallel sheets, favoring Fermi-surface-nesting driven CDWs in these materials.[23,26-28] Owing to a slight orthorhombicity in the crystal structure, with the *a*-axis smaller than the *c*-axis (e.g. a ≈ 4.312 Å and c ≈ 4.314 Å for $TbTe_3$ near the bulk $T_{cdw}$) [24], the unidirectional CDW preferentially forms along the *c*-axis at the higher-temperature CDW transition. However, it has been noted that there is only a moderate enhancement in the real and imaginary components of the Lindhard susceptibility at $q = 0.25c^*$ which is close to, but differs slightly from, the experimentally measured CDW wavevector of $q_{CDW} = 0.30c^*$, leading to questions as to whether

Fermi-surface nesting alone can fully account for the observed CDW in these compounds.[16,29] Recent inelastic x-ray scattering measurements on TbTe$_3$ find that at $T_{cdw}$ there is evidence for both a renormalization of the soft phonon mode toward zero energy and a peak in the linewidth for the soft phonon mode centered at $q = 0.3$**c***, consistent with the experimentally observed CDW wavevector.[29] In turn, these x-ray measurements, as well as Raman spectroscopy measurements conducted on ErTe$_3$ [30], suggest momentum-dependent electron-phonon coupling plays an important role in the formation of CDW states in the RTe$_3$ compounds. While a full understanding of the driving mechanism for CDW states in the RTe$_3$ compounds is still incomplete, it appears that both Fermi surface instabilities and momentum-dependent electron-phonon coupling may play a role.[31]

In this paper, we present results of temperature-dependent (298 K – 355 K) scanning tunneling microscopy (STM) studies on TbTe$_3$, which has a bulk $T_{cdw} \sim 336$ K below which a unidirectional CDW is established along the *c*-axis. In our measurements, we surprisingly detect spatially separated unidirectional CDW states along both the *a*- and *c*- crystal axes as well as regions where the perpendicular CDW states coexist. We propose that the multiple CDW orders established at the surface of TbTe$_3$ are driven by a strain field induced in a weakly-coupled Te surface layer through sample cleaving. Using bias-dependent measurements, we investigate the application of the 1-dimensional Peierls model to CDW states in TbTe$_3$. In addition, through temperature-dependent measurements, we find evidence for local CDW order above bulk $T_{cdw}$.

**Experimental Methods:**

TbTe$_3$ single crystals were grown using a self-flux technique described in detail elsewhere.[32] Our studies were conducted using an RHK PanScan STM operating in an ultrahigh vacuum (UHV) chamber with a base pressure $\sim 5 \times 10^{-10}$ Torr in a temperature range of 298 – 355 K. Atomically resolved STM tips were prepared by chemical etching of tungsten wire, followed by in-situ conditioning – annealing, followed by fine sharpening through electron bombardment. The TbTe$_3$ samples studied were cleaved in UHV by mechanically striking a cleave bar epoxied to the sample surface. Previous STM measurements [33,34] show that RTe$_3$ compounds cleave between the double Te planes such that the STM tip directly probes the exposed Te surface.

**Results:**

**a. Unidirectional CDW States:**

Figure 1b shows a topographic image of the surface of TbTe$_3$ acquired at 298 K. The exposed square Te lattice (often referred to as a Te square net) is observed. Superimposed upon the square lattice are parallel "stripes" due to an established unidirectional CDW state. The fast Fourier transform (FFT) of a typical topographic image (Figure 1c) shows three major components. First, there are four peaks (orange circles) due to the square Te surface layer with nearest neighbor spacing ~3 Å. Second, rotated 45°, are four peaks (blue circles) associated with a square lattice originating from the rare-earth block layer below. Density functional theory electronic-structure calculations indicate that the block layer current signal detected by STM is dominated by the rare earth ion [34], in our case Tb with nearest neighbor Tb spacing of ~4.3 Å. Third, the peaks in the FFT circled in yellow are due to the presence of the unidirectional CDW state.

The observation of a unidirectional CDW in TbTe$_3$ for temperatures just below $T_{\text{cdw}}$ is typically a good indicator of the *c*-axis of the crystal, given that a long-range CDW is established along the *c*-axis in the bulk. However, at different regions across the sample surface, and for the same crystal cleave, our measurements show extended regions (400 Å or larger) containing a unidirectional CDW in either of two perpendicular orientations. Furthermore, there is no general preference for the direction of the observed unidirectional CDW; it is equally likely that the CDW will be observed in either of the two directions. The establishment of unidirectional CDWs along two perpendicular crystal axes prevents us from unambiguously identifying the *a*- and *c*- crystal axes from our STM measurements. For this reason, in describing the CDW in our measurements, we often refer to the crystal axes in the *a-c* crystal plane as $a_1$ and $a_2$ instead of the traditional *a*- and *c*-axis designation. We find the CDW wavevectors along the two crystal axes are equivalent in magnitude: $q_1 = (0.30 \pm 0.01)a_1^*$ and $q_2 = (0.30 \pm 0.01)a_2^*$ where $a_1$* and $a_2$* differ by only ~0.002 Å$^{-1}$ at 300 K according to x-ray measurements [24].

Such an observation of unidirectional CDWs along both the *a*- and *c*- crystal axes is very surprising given the purely *c*-axis unidirectional CDW in the bulk. Bulk crystals can harbor stacking faults, corresponding to a 90° crystal misorientation during crystal growth, which could possibly account for the observation (i.e. successive cleaves of the same crystal could possibly reveal 90 degree rotation of the *a* and *c* axes). Based on individual STM measurements one cannot rule out such a scenario, although we note that x-ray measurements indicate, at minimum, that the vast majority of the sample is in a single orientation, whereas we observe perpendicular orientations of the unidirectional CDWs on multiple samples and sample cleaves with no general preferential direction to the observed orientation. Significantly, however, we observe regions of the sample where the two unidirectional CDWs coexist, providing compelling evidence that that unidirectional CDWs can indeed exist at the surface in both *a* and *c* directions. Furthermore, as we will discuss in greater detail in subsequent sections of this paper, these regions evolve continuously from *a*-axis oriented, to simultaneous order, to *c*-axis oriented CDWs upon translating across the surface of the sample, without encountering any step edges. Our measurements clearly indicate a relaxation at the surface of bulk constraints which would otherwise favor formation of the unidirectional CDW along the *c*-axis over the *a*-axis. This suggests that the surface Te-layer is only weakly coupled to the bulk below, a possibility on which we will expand in more detail in the discussion section.

**b. Wavevector Mixing and Tip-Condition Effects:**

Figure 1d shows a linecut through the FFT of the topography in the direction of the CDW, starting at the origin, and ending at $q_{atom}$, the peak associated with the block layer. In addition to $q_{atom}$, we label four peaks as "2", "3", "4", and "5" representing peaks at ~2/7, ~3/7, ~4/7, and ~5/7$q_{atom}$ respectively. These multiple peaks have been previously explained as originating from wavevector mixing of the true CDW signal with the atomic (Tb) block layer signal as well as mixing of their harmonics.[33,34] More specifically, wavevector mixing arises due to the asymmetric coupling between two sinusoidal signals. Given that the tunneling current has contributions from the top Te layer, in which the CDW state predominantly resides, as well as from the rare-earth block layer below, it is not surprising that these signals may become coupled through measurement setup conditions. As a consequence of this coupling, there arises a contribution to the total signal which resembles the product of two sinusoidal functions. Using standard trigonometric identities, this product can be re-expressed as the sum of two sinusoidal functions with periodicities given by the sum and difference of the original wavevectors.

While the STM studies of TbTe$_3$ by Fang et al. [33] and of the related CeTe$_3$ by Tomic et al. [34] observe the same four peaks in FFTs of topographic images, the two studies assumed differing fundamental CDW wavevectors, and consequently different wavevector mixing scenarios to explain the origin of the multiple peaks.

Fang et al. presents the following peak origins:

Peak 2: $q_{atom} - q_{cdw}$ ; Peak 3: $2q_{cdw} - q_{atom}$ ; Peak 4: $2q_{atom} - 2q_{cdw}$ ; Peak 5: $q_{cdw}$

Tomic et al. provides an alternative:

Peak 2: $q_{cdw}$ ; Peak 3: $q_{atom} - 2q_{cdw}$ ; Peak 4: $2q_{cdw}$ ; Peak 5: $q_{atom} - q_{cdw}$

The "true" CDW wavevectors identified in each scenario are equivalent up to a reciprocal lattice vector, so scattering measurements cannot distinguish between the two wavevector possibilities. However, if the wavevector mixing peaks arise as artifacts, purely due to measurement setup conditions as previously suggested [33-35], it should be possible to identify the true CDW wavevector. In particular, peak 4 in Tomic's wavevector-mixing scenario is purely the first harmonic of $q_{cdw}$. On the other hand, peak 4 in Fang's wavevector-mixing scenario results from the mixing of two signals: 1) the first harmonic of the atomic signal and 2) the first harmonic of the CDW signal. If one were to remove the first harmonic of the atomic signal, then any wavevector mixing peaks resulting from coupling to this true signal will also necessarily be removed. Specifically, if one were to remove the first harmonic atomic signal, and peak 4 remains, then peak 4 cannot be due to wavevector mixing.

It is possible to effectively do this by examining data where the tip is in a "good" but not "great" condition. In other words, the tip is good enough that, when acquiring a topography, we can identify the CDW, Te lattice, and rare-earth block layer peaks in the FFT. However, the tip is not "great" in that the harmonics of the atomic signals are not present (e.g. are indistinguishable from the noise level). The bottom plot in Figure 2a illustrates this situation where the first harmonic of the block peak ($2q_{atom}$) is clearly missing. However, all of the CDW and mixing peaks, including peak 4, are present as they are in the middle and top plots when the block harmonic signal is present. As a result, we conclude that the true CDW wavevector is $q_{cdw}$ ~2/7c*, not ~5/7c*. Consistent with this, the peak near ~1.3 in the bottom plot is present as expected for $q_{cdw}$ ~ 2/7c*, since this peak represents $q_{atom} + q_{cdw}$. However, the peak at ~1.4 is at the noise level, consistent with it disappearing. Since the ~1.4 peak results from $2q_{atom} - 2q_{cdw}$, its disappearance is to be expected when the harmonic signal $2q_{atom}$ is absent.

We emphasize that distinguishing between the two possible CDW wavevector mixing scenarios in the above fashion relies on the assumption that the wavevector mixing peaks originate purely from STM measurement setup conditions. However, it is well-established that there is a physical coupling of the Te planes to the neighboring block layers in the $RTe_3$ compounds.[36,37] Therefore, if the wavevector mixing peaks originate not from STM setup conditions as previously suggested, but rather from a physical coupling of the CDW to the block layer, then distinguishing between the two scenarios based on FFT peak differences, or lack-there-of, (arising from the presence or absence of the atomic harmonic signal) is not possible. However, if wavevector mixing peaks arise due to a physical coupling, this could potentially open the door to future investigations using STM to understand and characterize CDW coupling to the block layer.

We note that Fang et al. determined $q_{cdw}$ ~5/7c* based on their bias-dependent measurements.[33] Namely, at higher biases, they found the ~2/7c* peak intensity diminished very significantly whereas the ~5/7c* peak was noticeably prominent. We do not observe any such peak-intensity bias dependence in our measurements. Rather, we find that the peak intensities of the CDW and mixing peaks are dependent on the tip condition. As an example, Figure 2a shows that in one case (at top) the ~2/7c* peak is weaker than the ~5/7c* whereas the middle and bottom plots, the ~2/7 peak is stronger for measurements at the same bias.

Furthermore, Figure 2a illustrates that the relative peak intensities in the FFT, and consequently features observed in topographic images (Figures 2b, c, and d), are sensitive to slight differences in the tip condition. For certain tip conditions (the most typical case), the Te square surface net is most prominent in our images (Figure 2b). In other images, the rare-earth block layer is more prominent (Figure 2c). Furthermore, we have acquired topographic images where the tip condition is such that the sum of the Te and block signals produces a topography which has the appearance of dimerization (Figure 2d), a feature noted by Fang et al.[33] If we Fourier filter individual Te or block-layer peaks for the topography in Figure 2d, we see only the expected Te square lattice and the block square lattice respectively. However, when we filter to include both the Te and block peaks, we see the appearance of dimerization. We believe the previously-identified dimerization is due, in part, to a tip effect linked to the relative tunneling current components originating from the Te surface layer and the rare-earth block layer below, and is not evidence of a true surface dimerization. We provide more details regarding this dimerization appearance in the discussion section.

Since we have highlighted the effects due to slight changes in the tip condition, it is worthwhile to emphasize what does and what does not change with tip condition. While the relative intensity of peaks in the FFT are linked to tip condition, and hence the topographical features which might be immediately apparent visually, the CDW, wavevector mixing, and lattice wavevectors associated with the peaks do not change. As a result, other than exploiting the tip condition to aid in distinguishing between a 2/7 and 5/7 CDW wavevector, we emphasize that the results detailed in this manuscript are independent of tip condition.

**c. Bias-Dependent Measurements:**

While the $RTe_3$ compounds were initially believed to be prototypical Fermi-surface-nesting-driven CDW compounds, more recent studies indicate the importance of momentum-dependent electron-phonon coupling in the establishment of CDW states in these materials. Bias-dependent STM measurements have the potential to give insight into the origin of CDW states. Namely, in the case of a CDW state described within the 1-dimensional Peierls model, the electron and hole components of the CDW are expected to be spatially out of phase.[16,19,38] Such a model has been applied to quasi 1-dimensional systems such as the blue bronzes and $NbSe_3$ where ARPES and band structure calculations determine that there are substantial parallel components of their Fermi surfaces.[12,13,39,40] In $NbSe_3$, the expected bias-dependent contrast inversion in the CDW state has been detected by STM [41] despite imperfect nesting [40]. To our knowledge, similar measurements have not been conducted on the blue bronzes.

Quasi 2-dimensional materials such as $K_{0.9}Mo_6O_{17}$ and $\eta$-$Mo_4O_{11}$ have been shown to have a "hidden" one-dimensional nature [42-45], motivating the possibility of applying a 1-dimensional Peierls model to understand CDW states in these systems. Whereas the Fermi surface of $K_{0.9}Mo_6O_{17}$ leads to "extremely good nesting" conditions [43], the Fermi surface of $\eta$-$Mo_4O_{11}$ allows for only imperfect nesting [44]. In each case, STM has directly imaged bias-dependent contrast inversion in the CDW state, which has been used to support the case for CDW states described by the 1-dimensional Peierls model arising in a 2-dimensional system.[46,47] Similar to $\eta$-Mo$_4$O$_{11}$, tight-binding modeling of the Te layer in the RTe$_3$ compounds reveals a 1-dimensional character to the material due to the anisotropy of the $p_x$ and $p_z$ orbitals within the Te plane.[28] In the idealized model, the 1-dimensional chains separately formed by the $p_x$ and $p_z$ orbitals would lead to a near-perfect nesting-driven CDW state. However, coupling between the chains leads to an imperfect nesting scenario [48], similar to $\eta$-Mo$_4$O$_{11}$.

We acquired bias-dependent measurements to determine whether there is a phase shift in the CDW state imaged at positive and negative biases near the Fermi energy. Figure 3a and 3b show topographic images taken over the same 154 Å square region at +100 mV and -100 mV respectively. We Fourier filtered the two images to visually enhance three common "dark spot" surface features which we have circle with white ovals (Figures 3c and 3d). We emphasize that the white ovals are of exactly the same size and extend over identical regions in the two images. We then Fourier filtered the images such that only the CDW contribution is included in the images (Figures 3e and 3f). We note that the CDW maxima and minima appear to align with each another. To examine this more closely, we enhanced the CDW signals in Figures 3e and 3f by a factor of 15 and added it to the original filtered topographies (Figures 3c and 3d) resulting in the images seen in Figures 3g and 3h. In the images appearing in figures 3g and 3h, using as reference the common three white ovals, it is clear there is no phase shift between the CDW maxima and minima seen at +100 mV and -100 mV. In contrast to other 2-dimensional CDW compounds with a 1-dimensional character, we do not observe bias-dependent contrast inversion of the CDW state in our measurements. Apparently the simple 1-dimensional Peierls model does not apply to TbTe$_3$, possibly reflecting contributions to the density of states from reconstructed (metallic) parts of the quasi 2-dimensional Fermi surface, and perhaps suggestive that factors beyond Fermi surface nesting might play a significant role in the CDW formation.

### d. Temperature Dependence:

While the bulk CDW transition in TbTe$_3$ has been determined by resistivity measurements to occur at $T_{cdw}$ ~336 K, x-ray measurements have observed superlattice peaks up to 363 K.[24] These CDW-associated superlattice peaks have been attributed to CDW fluctuations. The presence of such fluctuations indicates the possibility of localized CDW order above the bulk transition temperature. Such localized CDW order above bulk $T_{cdw}$ has been directly visualized by STM in NbSe$_2$; defects act as nucleation sites from which the long range CDW forms with decreasing temperature near $T_{cdw}$.[18]

We conducted temperature-dependent measurements from 298 K, below the bulk $T_{cdw}$, to 355 K, above $T_{cdw}$. At all temperatures within this range (taken at ~5 K increments, or in ~1 K increments between 325 K – 347 K), we have been able to detect a unidirectional CDW state. Figure 4a shows a 90 Å square topographic image acquired at 355 K in which a unidirectional CDW state is clearly observed. A linecut through the FFT in the direction of the CDW shows the standard four associated CDW and wavemixing peaks (Figure 4b). In our measurements at 345 K, we find regions where CDW order is present and others where it is absent, indicating strong variations in the CDW at this temperature. Such variations are particularly obvious in our measurements at 339 K. Figure 4c and 4d show a 240 Å x 200 Å topographic region (raw image and Fourier filtered to enhance CDW respectively) where there are variations in the unidirectional CDW along the $a_1$-axis. Specifically, the CDW state in regions in the left half of the image appears weak or even absent. To emphasize these spatial variations in the CDW state, we took an FFT of the left half of the image and separately an FFT of the right half of the image (Figure 4e). We subtracted the approximately exponential background from the linecut through the FFTs to allow for quantitative comparison of the CDW peaks. We find that the ~2/7 $a_1$* peak is ~3 times larger for the right side of the image compared to the left.

Our measurements detect localized CDW order and spatial variations in that order at temperatures above the bulk transition. We propose that instead of nucleating from surface or subsurface defects, such local formation of CDW states is attributed to weak coupling of the Te layer to the bulk, coupled with strain variations across the surface layer, possibly leading to spatially-varying and locally-associated surface $T_{cdw}$s. We will provide support for this surface decoupling in the discussion session, as well as motivate the origin of the strain variations. Finally, in our measurements, we find no temperature-dependent preferential CDW orientation ($a_1$ versus $a_2$ axis) and no temperature evolution in the magnitudes of the measured CDW wavevectors $q_1$ and $q_2$.

### e. Coexisting CDW States:

We have acquired data in which the FFT indicates the presence of CDWs along both the $a_1$ and $a_2$-axes (hence $a$-axis and $c$-axis) within a ~500 Å square region (Figure 5a). Due to noise at lower wavevectors, we focus on the four wavevector mixing peaks around the block layer signal, $\vec{q}_{atom} \pm \vec{q}_1$ and $\vec{q}_{atom} \pm \vec{q}_2$, which are circled in yellow in Figure 5a, and are shown in detail in Figure 5b. The small yellow circles in Figure 5b indicate the presence of a CDW along the $a_1$ axis and the red circles indicate the presence of a CDW along the $a_2$ axis.

To acquire a more local view, we broke the ~500 Å square image into smaller regions. Figure 5c shows a 150 Å square image in which the CDWs along both the $a_1$ and $a_2$ axes spatially coexist. We then examined 150 Å square regions directly to the left and to the right of this "middle" region where the two CDW states coexist. The "left" region shows an $a_1$-axis-dominated CDW and the "right" region shows an $a_2$-axis-dominated CDW (Figures 5d and 5e). The "middle" 150 Å region appears to not only surprisingly represent a region where two perpendicular CDW states coexist, but appears to be a transition region between regions where one of the unidirectional CDWs is more dominant. This possibly smooth transition from an $a_1$-axis-dominated CDW to an $a_2$-axis-dominated CDW region is clearly indicated by Figure 5f.

Such coexistence of perpendicular CDW states in TbTe$_3$ appears in the bulk only below ~41 K.[25] In the bulk, the $a$-axis CDW has wavevector of ~(1-.68)$a$* = 0.32$a$*, compared to the $c$-axis CDW wavevector of ~(1-0.71) = 0.29$c$* at 10 K.[25] The coexisting CDWs and slight differences in their wavevector magnitudes were first detected in STM measurements by Fang et al., and they noted identical wavevector differences between the two CDWs. On the other hand, in our measurements conducted near room temperature, we find $q_{1,2} \approx 0.30 a^*_{1,2}$ (extracted from the FFT of the ~500 Å square image), which is consistent with the wavevector magnitudes when only a single unidirectional CDW is present in a given location at the surface.

It is very surprising to observe coexisting CDWs at temperatures ~7.5 times higher than the low-temperature CDW transition in TbTe$_3$. Furthermore, the wavevector magnitudes for the two coexisting CDWs are equivalent near room temperature, but have a noted difference at low temperature, even by STM. Such a difference may be accounted for by temperature-dependent differences in the coupling of surface states to the bulk. Namely, at the elevated temperatures at which we are conducting our experiments, the surface Te-layer is more weakly coupled to the bulk than at lower temperatures.

Finally, we note that the spatially coexisting CDW orders we observe differ from the bidirectional checkerboard state theoretically detailed by Yao et al.[28] Yao et al. suggest that checkerboard order competes with unidirectional stripe CDW order in the RTe$_3$ compounds.

Specifically, in their analysis, the Fermi surface nesting vectors leading to bidirectional order are rotated relative to the individual nesting vector possibilities leading to unidirectional CDW order. In our measurements of TbTe$_3$, we find the CDW wavevectors for the bidirectional order are identical to those of the unidirectional states. This agreement points towards a non-nesting scenario driving CDW order in this material.

**Discussion:**

Our STM measurements indicate differences in the CDW states at the surface compared to that of the bulk. In particular, at the surface, we observe spatially-separated as well as spatially-coexisting unidirectional CDWs along the both the *a*- and *c*- crystal axes at room temperature and above, whereas only a *c*-axis unidirectional CDW is detected in the bulk in this temperature range. Such differences suggest that the surface Te layer is only weakly coupled to the bulk below. To investigate whether there is evidence that the surface Te layer is decoupling from the bulk, we performed detailed analysis of topographic images to determine whether there is evidence for a shift in the Te ions from their expected lattice sites. Specifically, because the tunneling current contains a contribution from the surface Te layer and a contribution from the block layer originating from the Tb ions, it is possible to use Fourier filtering to determine the crystal structure specific to each layer as well as the relative positions of the two sets of ions. Figure 6a shows the bulk crystal lattice locations of the Te ions (in red) from the Te layer and nearest Tb ions (in blue) from the block layer below, the two ions which dominate the tunneling current signal, as seen projected onto the *a-c* plane.

We present analysis on a single ~120 Å square topographic image from which a 90 Å square appears in Figure 1b. We Fourier filtered the 120 Å square topographic image so as to include only the Te and block layer lattice contributions. Because the topographic image is Te-dominated, the block layer signal was enhanced by a factor of ten during the filtering process so that the Te and block layer signals are of comparable size. Figures 6b, c, and d show three 25 Å square regions cropped from the resulting filtered image taken from within the bottom, middle, and top thirds of the image respectively. Visually the three images differ significantly, an indication that the Te ions may be in differing locations relative to the Tb ions in each of the three regions. Separately filtering the Te and Tb signals allows for a determination of the Te and Tb ion locations for each of the three regions. Figures 6e, f, and g show the locations of each ion, Te in red and Tb in blue, for each region overlaid on the 25 Å images. To further elucidate the ion locations, Figures 6h, i, and j show the ion locations for the three regions. There appears to be a progression, where, at the bottom of the image, the Te ions are in their expected locations based on the bulk crystal structure, are shifted noticeably in the middle region, and even more so in the top region. These shifts cannot be explained as due to the periodic lattice distortion induced by the CDW state as the shifts seen in the middle and top regions occur predominantly along the $a_2$ axis whereas the local CDW state is induced along the $a_1$ axis.

This analysis provides strong evidence for a decoupling of the Te surface layer from the rare-earth block layer below. Such a decoupling is not entirely surprising since it has been noted previously [21] that the Te-Te bond length in the Te layer of the bulk which is ~3.1 Å is considerably longer than the typical Te-Te covalent bonding length of 2.76 Å [49], leading to a structural instability. As a consequence, upon cleaving, the surface Te layer is no longer constrained from above, as it would be in the bulk, allowing the surface to relax to a more stable configuration. In addition, the cleaving process itself can induce strain variations across the weakly coupled surface layer, possibly leading to surface corrugations. As a consequence, we propose that these local strain variations affect the specifics of electron-phonon coupling in the surface Te layer, which, in turn, determine the axial direction of the locally established unidirectional CDW, or in rare cases, establishes coexisting bidirectional order. Similar strain variations in the surface of $NbSe_2$ due to subsurface defects were proposed to drive local CDW order from tri-directional to unidirectional, whereas only the tri-directional ordering is reported in the bulk.[17] In addition, a very recent study combining x-ray and density functional theory calculations shows that anisotropic lattice strain energy plays a critical role in establishing the unidirectional CDW state along the *c*-axis in the bulk of $TbTe_3$.[31] The sensitivity of the CDW direction in the bulk to lattice strain illustrates how strain variations in the Te surface layer could drive the two perpendicular CDW orders observed in our measurements.

Additional analysis of the 120 Å square region is consistent with this interpretation. The lattice parameters $a_1$ and $a_2$ are the expected 4.31 ± 0.01 Å and 4.31 ± 0.03 Å respectively based on the Tb ion locations. Using the Te ion locations to calculate the lattice parameters, we find $a_1$ = 4.28 ± 0.37 Å and $a_2$ = 4.25 ± 0.06 Å. These smaller lattice-parameter averages, based on the Te ion locations, are consistent with an overall Te lattice relaxation. The larger standard deviations can be accounted for as strain-induced local lattice variations. We note that there is an expected Te lattice parameter variation of ~3% along the CDW axis due to the CDW modulation [23] corresponding to ~ 0.13 Å. The Te lattice variations along the CDW axis, $a_1$, are considerably larger by about a factor of 3.

Finally, our analysis also indicates the origin of the appearance of dimerization, as previously noted by Fang et al. [33] in their STM measurements and as seen in some of our acquired topographies such as Figure 2d. This dimerization, as elucidated by Figure 6g, is the result of two factors: 1) a shift of the Te ions from their expected location and 2) a tip condition such that the components of tunneling current originating from the Te and block layers are similar in magnitude. While this dimerization does not represent an additional true broken symmetry at the surface, it is an important indicator that the surface Te layer is decoupling from the block layer below.

**Conclusion:**

Our studies motivate the role of strain in driving the observed multiple CDW orderings at the surface of $TbTe_3$, most likely by modifying local electron-phonon coupling within the surface layer. The interplay of lattice strain anisotropy and electronic properties has been shown to be important in establishing the CDW states in the bulk of the $RTe_3$ compounds. Our measurements suggest the same is true at the surface. Using STM to simultaneously study local strain and local electronic properties has the potential to provide nanometer insight into this interplay.

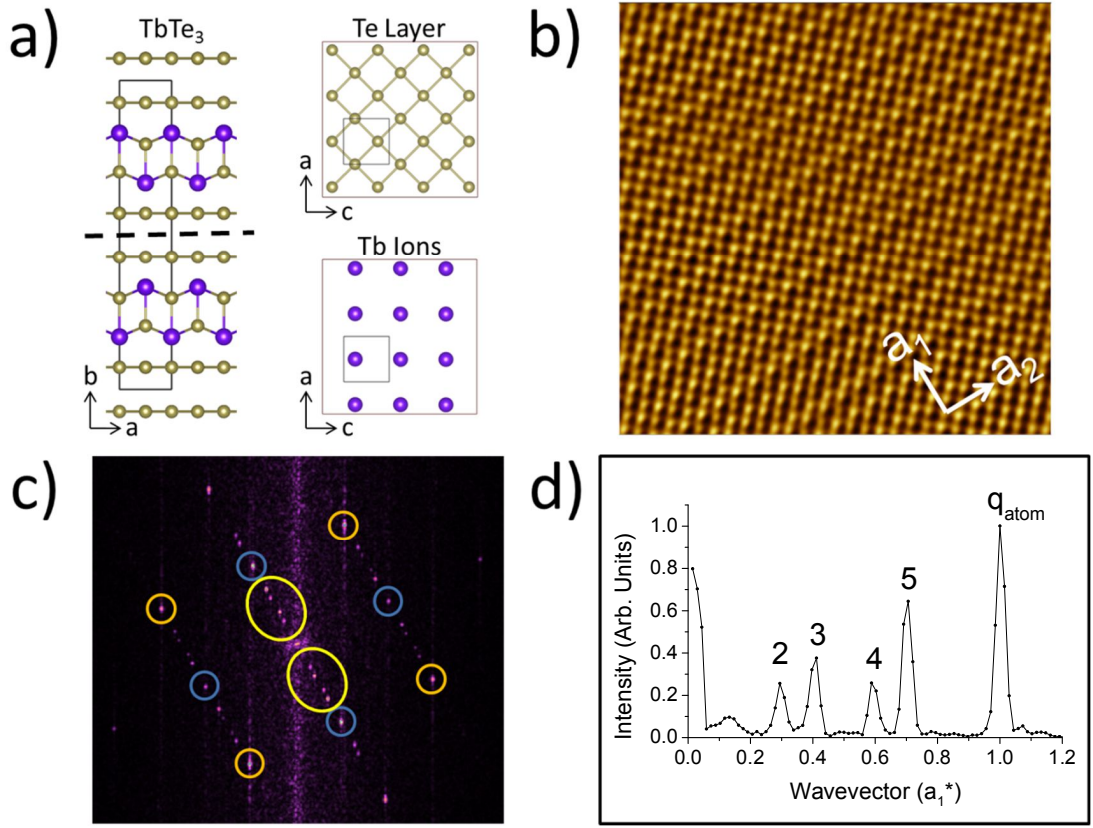

**Figure 1**: a) At left, the crystal structure for TbTe$_3$ with a black rectangle outlining the unit cell. The dotted line indicates the *a-c* cleave plane between the double Te layers. At right, the square lattice of the Te layer which is exposed by cleaving as well as the locations of the closest Tb ions in the rare-earth block layer directly below. The unit cell is again shown in the structures at right for reference. The crystal structures were constructed using Vesta software [50]. b) Topographic image taken over a 90 Å square region at I = 65 pA, V$_{Sample}$ = -350 mV. The Te-square lattice of the exposed surface can be clearly seen as well as superimposed "stripes" associated with a unidirectional CDW state along the $a_1$ crystal axis. We use $a_1$ and $a_2$ to denote the in-plane crystal axes for our measurements since we observe unidirectional CDW order along both axes. This prevents us from unambiguously identifying the *a*- and *c*-crystal axes. c) FFT of a typical topographic image. Orange circles identify the wavevectors associated with Te-square lattice. Blue circles identify the wavevectors associated with the subsurface rare-earth block layer (Tb ions). The yellow ovals enclose peaks in the FFT which are associated with the unidirectional CDW as well as those arising from mixing between the CDW wavevector and block atomic wavevectors. d) Linecut through the FFT, beginning at the origin, in the direction of the CDW, and ending just past $q_{atom}$ associated with the block layer.

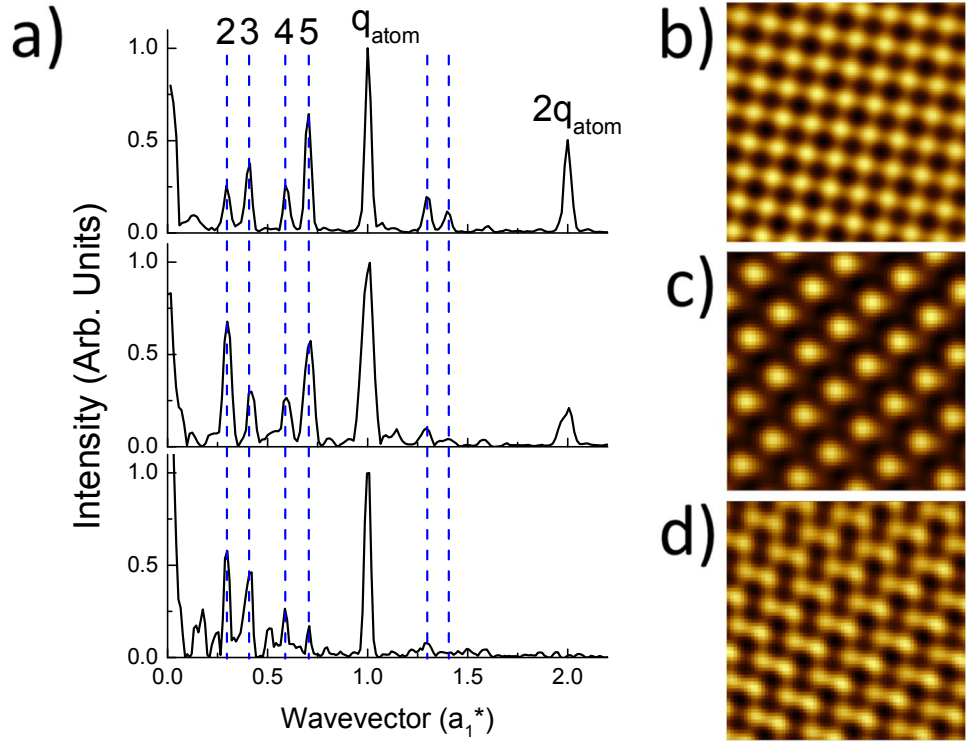

**Figure 2**: a) Linecuts through FFTs in the direction of the CDW. The harmonic of the block signal, $2q_{atom}$, is present in the top in middle plots but absent in the bottom plot. Peak 4 is present in all three cases indicating that peak is not the result of wavevector mixing involving the block harmonic, as the harmonic is absent in the bottom plot but peak 4 is still present. Slightly different tip conditions lead to variations in the relative intensities of the peaks in the FFT. Each of the FFTs, from which the linecuts are extracted, were taken on 400 Å square topographic images acquired with the same settings: $I = 50$ pA and $V_{sample} = +150$ mV with the same number of pixels and at temperatures at least 10 K below the bulk $T_{cdw}$. b),c),d) 25 Å square topographies which have been Fourier filtered to include only contributions from the Te net and rare-earth block signals. Small differences in tip conditions lead to a Te-dominated topography (b), a rare-earth block-dominated topography (c), and a topography which has the appearance of surface dimerization (d).

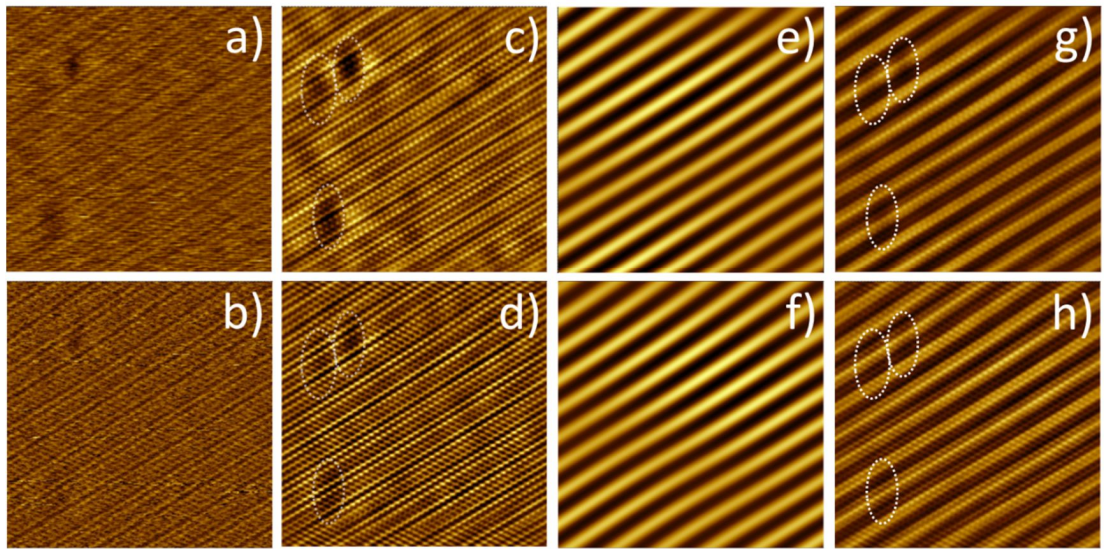

**Figure 3**: a) and b): 154 Å square topographies taken over the same exact location with I = 30 pA and $V_{Sample}$ = +100 mV (for a) and $V_{Sample}$ = -100 mV (for b). c) and d): Images in a) and b) Fourier filtered to include the Te, block, and CDW/wavevector mixing signals as well as low wavevector signals. The three dark regions visible in a) and b) are enhanced through the filtering and circled with white ovals. The white ovals extend over identical regions in the two images. e) and f): Using Fourier filtering, only the ~$2/7a_1$ CDW for +100 mV (in e) and -100 mV (in f) is shown. There is no phase shift evident in the CDW as imaged at positive or negative biases. g) and h): The CDW signals in e) and f) are enhanced by 15 times and added to the filtered image in c) and d) respectively allowing for identification of the CDW maxima and minima relative to the three ovals for +100 mV (in g) and -100 mV (in h). Using the ovals as a guide, these images clearly indicate that the CDW at positive and negative biases are in phase.

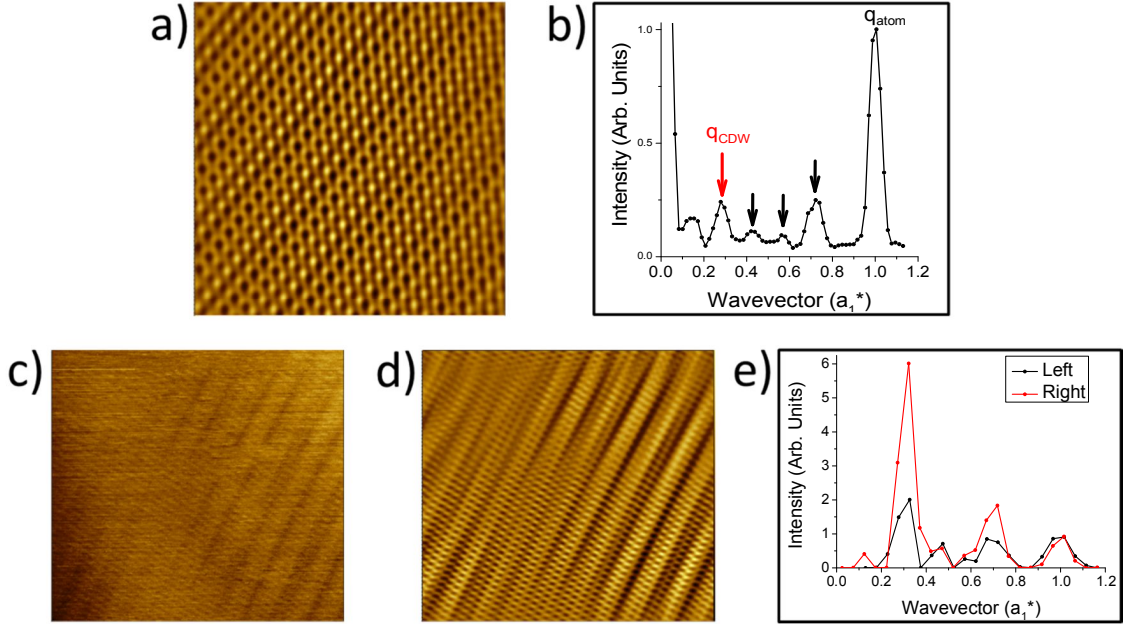

**Figure 4**: a) 90 Å square topography taken at 355 K with $V_{sample}$ = +200 mV and I = 40 pA. The image was Fourier filtered to include the surface Te lattice, block layer, and CDW/wavevector mixing signals. b) Linecut through the FFT (of the raw topographic image for a) beginning at the origin, in the direction of the CDW, ending at the $q_{atom}$ associated with the block layer. The standard 4 CDW/wavevector mixing peaks are observed. c) A 240 Å x 200 Å topography acquired at 339 K showing variations in the unidirectional CDW state across the region. $V_{sample}$ = -200 mV and I = 70 pA d) The image in c) Fourier filtered to include Te net, block layer, and CDW/wavevector mixing signals to visibly emphasize local variations in the CDW state. e) A comparison of the linecuts through the FFTs separately for the left and right halves of image c). The $\sim 2/7 a_1^*$ CDW peak is ~3 larger in the right half as compared to that of the left illustrating strong variation in the $a_1$-axis CDW in this region.

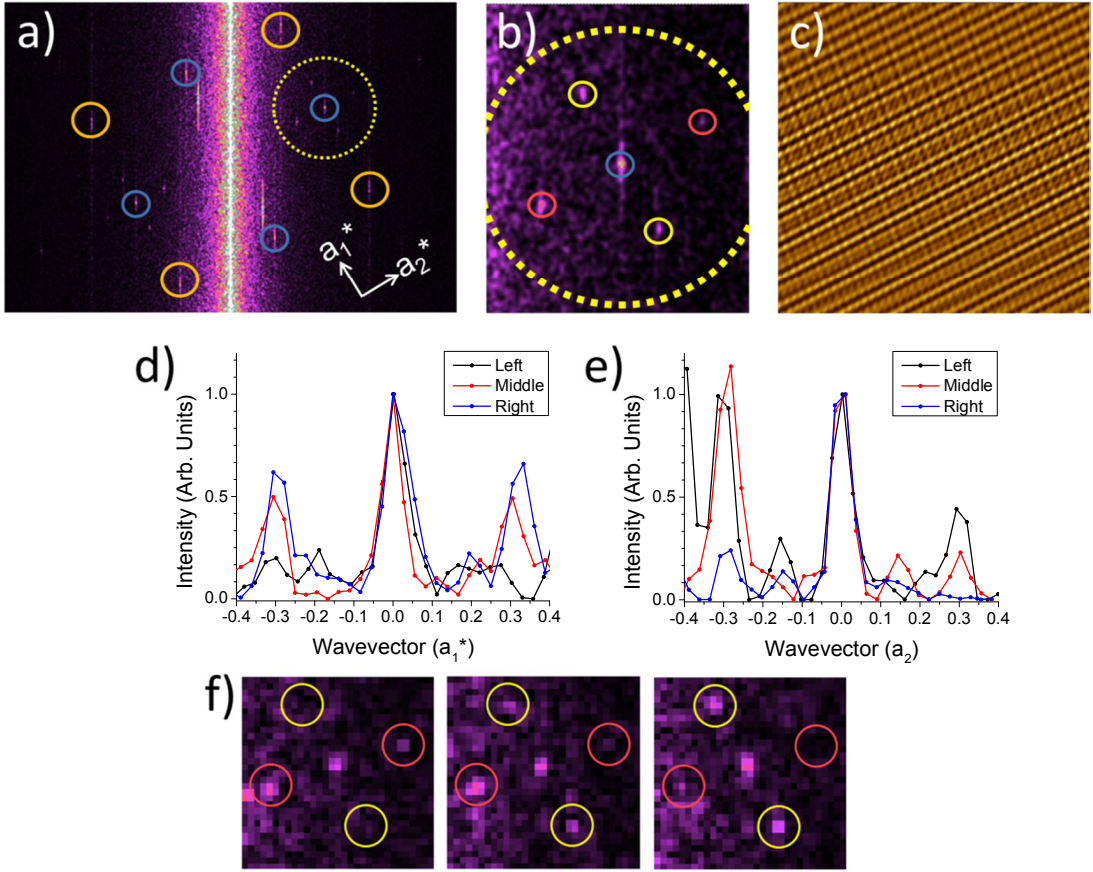

**Figure 5**: a) FFT of a ~500 Å square topography taken at 309 K with $V_{sample}$ = +150 mV and I = 50 pA. Standard lattice peaks are circled in orange (Te net) and blue (Block layer). In addition to these peaks, we observe peaks associated with two perpendicular CDWs within this field of view. Due to noise at low wavevectors, these CDW peaks are most clearly seen in the square pattern surrounding the block peak (circled in yellow). b) The region within the yellow circled area of the FFT in a) is enlarged to evince the block layer and CDW peaks. The peaks associated with the CDWs along the $a_1$ axis and $a_2$ axis are circled in yellow and red respectively. c) A 150 Å square region cropped from the larger topographic image where the two CDWs spatially coexist with one another. The image was Fourier filtered to reduce noise which obscured some of the topographic features. d) and e): A comparison of CDW-associated peak intensities associated with the $a_1$-axis (in d) and $a_2$-axis (in e) CDWs across neighboring 150 Å square regions. The rare-earth block layer peak is centered at 0 and the CDW-associated peaks are seen near ±0.3. There is a progression from an $a_2$-axis dominated CDW at left to an $a_1$-axis dominated CDW at right. f) This progression is easily seen visually as the $a_1$-axis CDW peaks (in yellow) around the block-signal in the FFT become more intense from left to right while the $a_2$-axis CDW peaks (in red) become less intense. The three FFTs are zoomed similar to what was done in b) with the block layer signal centered, and share a common color scale.

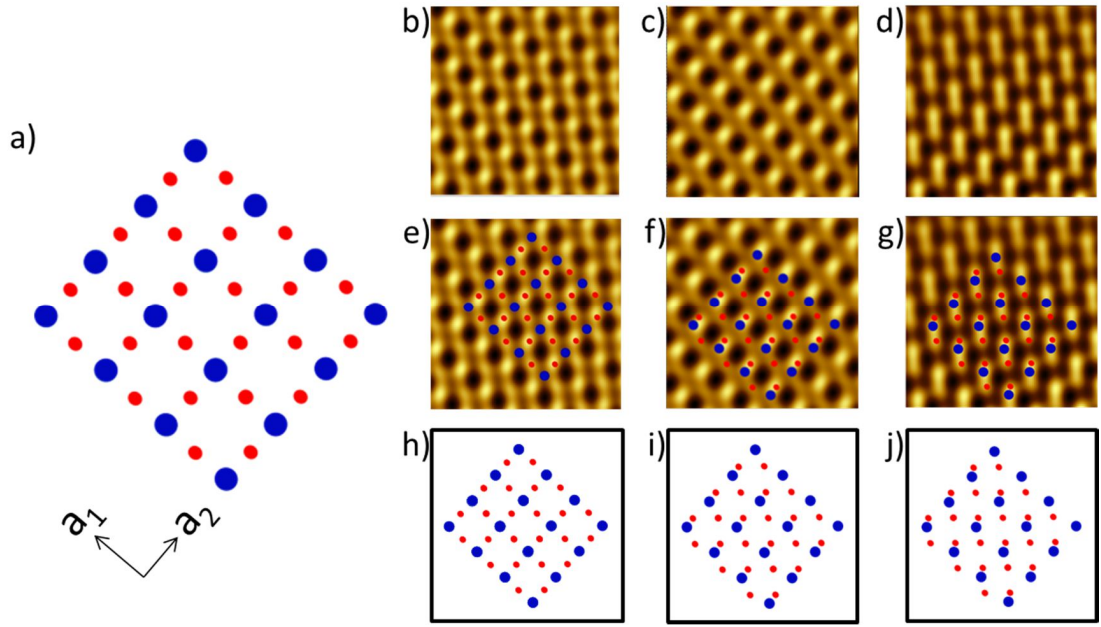

**Figure 6**: a) Relative bulk positions of Te ions (red) in the Te layer and nearest Tb ions (blue) from the rare-earth block layer directly below, projected onto the *a-c*. The axes are labeled as $a_1$ and $a_2$ since we are unable to distinguish the *a*- and *c*-axes in our measurements. b) c) and d): 25 Å square images cropped respectively from the bottom, middle, and top thirds of a Fourier-filtered 120 Å square image. The Fourier-filtered image only includes contributions from the Te and block-layer (Tb) structural signals. The block-layer signal was enhanced by a factor of 10 such that the Te and block layer signals are comparable. The differing images suggest differing relative locations for the Tb and Te ions in the three regions. e) f) and g): The specific lattice positions of the Te and Tb ions superimposed on the bottom, middle, and top images. For clarity, the Te and Tb ion positions for the bottom, middle, and top images are shown separately in h), i), and j) respectively. The shifts in the relative ion positions suggest a decoupling of the Te-layer from the rare-earth layer below.

**Acknowledgements:**
We thank Eric Hudson and Ming Yi for their comments on the manuscript. Samples used in this study were grown at Stanford University, supported by the Department of Energy, Office of Basic Energy Sciences under contract DE-AC02-76SF00515. This work is also supported by Clark University (faculty development grants and university and physics department research student support).

L.F. and A.M.K. contributed equally to the manuscript.